\documentclass[a4paper, apl, aip, runinaddress, reprint]{revtex4-1}
\usepackage{color}
\usepackage{graphicx}
\usepackage{amsfonts}
\usepackage{amsmath}
\usepackage{amssymb}
\usepackage{textcomp}
\usepackage{verbatim}
\usepackage{ulem}
\usepackage{epstopdf}
\usepackage{multirow}
\usepackage{eqnarray}
\usepackage{setspace} 
\usepackage{bigdelim,
			booktabs}

\begin{document}

\title{Spin-wave logic devices based on isotropic forward volume magneto-static waves}

\author{S.~Klingler}
\email{stefan.klingler@wmi.badw-muenchen.de}
\altaffiliation{Current Address: Walther-Mei{\ss}ner-Institut, Bayerische Akademie der Wissenschaften, 85748 Garching, Germany}

\author{P.~Pirro}
\altaffiliation{Current address: Institut Jean Lamour, Universit\'{e} de Lorraine, 54011 Nancy, France}

\author{T.~Br\"acher}
\author{B.~Leven}
\author{B. Hillebrands}
\author{A.~V.~Chumak}
\affiliation{Fachbereich Physik and Landesforschungszentrum OPTIMAS, Technische Universit\"at
Kaiserslautern, 67663 Kaiserslautern, Germany}

\date{\today}

\begin{abstract}

We propose the utilization of isotropic forward volume magneto-static spin waves in modern wave-based logic devices and suggest a concrete design for a spin-wave majority gate operating with these waves. We demonstrate by numerical simulations that the proposed out-of-plane magnetized majority gate overcomes the limitations of anisotropic in-plane magnetized majority gates due to the high spin-wave transmission through the gate, which enables a reduced energy consumption of these devices. Moreover, the functionality of the out-of-plane majority gate is increased due to the lack of parasitic generation of short-wavelength exchange spin waves.

\end{abstract}

\maketitle
Steadily advancing progress in modern information technology pushes the computation capabilities of common silicon-based digital logics towards fundamental quantum limits. This raises the problem of how to overcome these inherent problems. Furthermore, issues such as waste heat production demand for new approaches.\cite{ITRS2013,Galatsis2009} A potential alternative is wave-based computing. In particular, logic elements and devices based on collective excitations of the magnetization in the solid state - spin waves and their quanta magnons - have attracted attention in the recent years.\cite{Stamps2014, Schneider2008,Khitun2010,Bracher2013,Klingler2014a} In a spin-wave device the information can be encoded into the phase or amplitude of the wave, and the information can be processed by employing interference between different waves as well as nonlinear interactions.
This field is still in its infant state, but recent progress has demonstrated its large potential.\cite{Chumak2012, Chumak2014}

A cornerstone of wave-based logic elements is the majority gate, since it allows for a simple implementation of complex logic circuits and the processing of several boolean operations with a single gate structure.\cite{Khitun2008, Klingler2014a}
In a previous study, we introduced the design of an in-plane magnetized spin-wave majority gate.\cite{Klingler2014a} It consists of three input waveguides where spin waves are excited, a symmetric spin-wave combiner which merges the different input waveguides, and an output waveguide where a spin wave propagates with the same phase as the majority of the input waves. In the combiner region scattering processes into higher-order dipolar spin-wave modes and small-wavelength exchange spin waves occur due to the broken translational symmetry of the waveguide system and the anisotropic dispersion relation of the spin waves in this magnetization configuration. We showed, that parasitic scattering processes into higher dipolar modes can be suppressed with a suitable waveguide geometry. But still, the output signal is influenced by exchange spin waves.\cite{Klingler2014a} 

To overcome these limitations the use of isotropic forward volume magneto-static spin waves (FVMSW) is an interesting option, which thus implies the necessity of a new suitable majority gate design. Here, we demonstrate the functionality of an out-of-plane magnetized spin-wave majority gate which operates with isotropic forward volume magneto-static spin waves (FVMSW) and, thus, overcomes the limitations of the in-plane magnetized gates. We employ numerical simulations to prove the characteristics of the gate and we find a high spin-wave transmission through the gate of up to 64\,\%, which is about three times larger than for the in-plane magnetized gate.\cite{Klingler2014a} 

The simulations are performed for Yttrium-Iron-Garnet-(YIG)-structures with a thickness of 100\,nm.\cite{Klingler2014, Pirro2014} For this purpose, the following material parameters have been used: a saturation magnetization of $M_\mathrm{s}=140$\,kA/m, an exchange constant of $A=3.5$\,pJ/m and a Gilbert damping of $\alpha=5\cdot 10^{-4}$. Due to the small Gilbert damping parameter of YIG, spin-wave propagation distances in the millimeter range are observed.\cite{Serga2010} This is larger than the size of conventional microstructures and, thus, makes it a very suitable material for the construction of complex magnonic networks.\cite{Chumak2014} Furthermore, the relatively small saturation magnetization of YIG allows to switch the magnetization out-of-plane with moderate external fields ($\mu_0 H \gtrsim 180$\,mT), which can be easily applied by, for instance, a permanent magnet. 

\begin{figure}[t]%
\begin{center}%
\scalebox{1}{\includegraphics[width=\linewidth,clip]{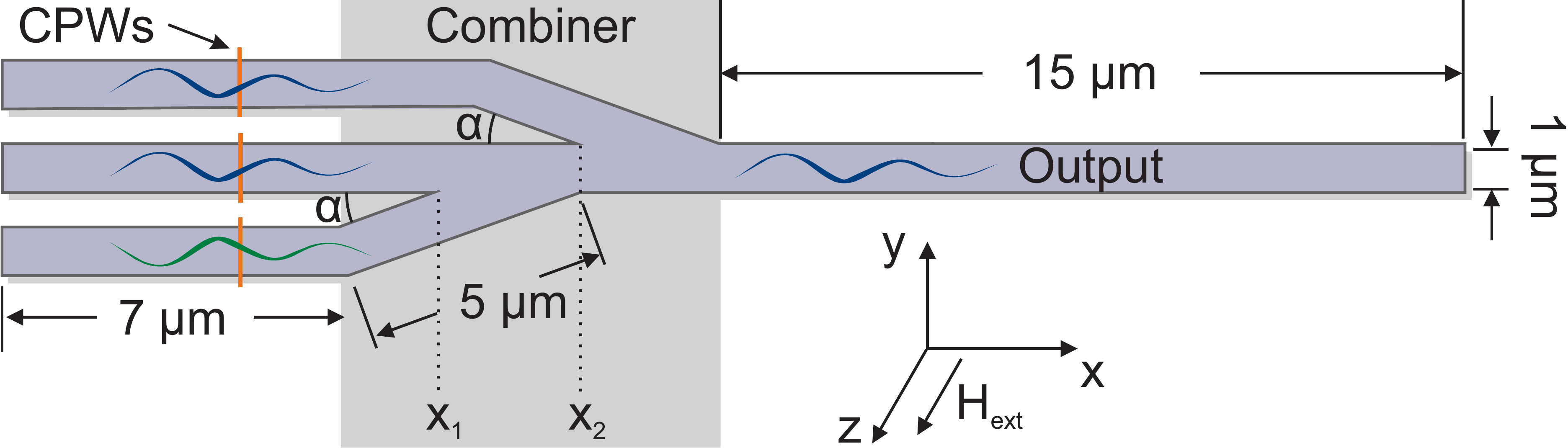}}%
\end{center}%
\caption{\label{sat_mask} 
Geometry of the majority gate. Spin waves are excited with CPW's in the input waveguides. The input waveguides are bent under an angle $\alpha$ and merged with the center input arm at positions $x_1$ and $x_2$ in the combiner region, where the spin waves interfere. The amplitude and phase of the spin wave in the output waveguide are then determined by the majority phase of the input spin waves.}%
\end{figure}%
	
In Fig.~\ref{sat_mask} the design of an out-of-plane magnetized majority gate is shown. All waveguides have a width of 1\,$\mu$m, so that the possibility of a practical realization of the gate structures is ensured.\cite{Chumak2014,Pirro2014,Hahn2014}
The majority gate structure consists of three parallel input waveguides, where the spin waves are excited and the information is encoded into the spin-wave phase. In the combiner the input waveguides are bent under an angle $\alpha$ towards the center waveguide, to overlay the spin waves and allow for their interference with each other. The bent parts have a length of 5\,$\mu$m and merge with the center waveguide at positions $x_1$ and $x_2$, respectively. With this asymmetric design the spin waves can be forced to propagate into the direction of the connected output waveguide, which guarantees a high energy transmission. In the output waveguide the output spin wave propagates with the same phase as the majority of the input waves.

The majority gate structure is investigated with numerical simulations using MuMax2.\cite{Vansteenkiste2014} The software performs the simulations on the graphics cards of the computer and, thus, allows for a highly parallelized calculation of the spin-wave dynamics in a mesoscopic magnetic system with a resolution in the order of the exchange length. Spin-wave reflections at the end of the waveguides are avoided by increasing the damping by a factor of 300 over the last 4\,$\mu$m in the $x$-direction. The external field was applied parallel to the $z$-axis with a strength of $\mu_0 H=200$\,mT. The cell size in the simulated area was chosen to be $14\times 8 \times 100$\,nm$^3$, so that the resolution is in the order of the exchange length ($\lambda_\mathrm{ex}=18$\,nm)\cite{Abo2013} of YIG and smaller than the spin-wave wave length in the microstructures. In $z$-direction the usage of only one cell is justified since no perpendicular standing spin-wave modes are excited at the working frequency of $f=1.5$\,GHz. To excite the spin waves, individual coplanar waveguides (CPW) were modeled for each input arm, to suppress dynamic magnetic fields outside of the excitation area. The center conductor of the CPW has a width of 400\,nm and a height of 250\,nm. The ground plates have widths of 200\,nm and heights of 250\,nm. The center-to-center distance between conductor and ground plate is 400\,nm. The excitation field is then calculated using Biot-Savart's law for an AC-current with an amplitude of 0.1\,mA in the conductor and -0.05\,mA in the ground plates and a frequency of 1.5\,GHz.
After an excitation time of 100\,ns the spin-wave amplitude reached its steady state. Subsequently, the magnetization distribution in the waveguide system was saved with a time resolution of 4\,ps.

\begin{figure}[t!]%
\begin{center}%
\scalebox{1}{\includegraphics[width=\linewidth,clip]{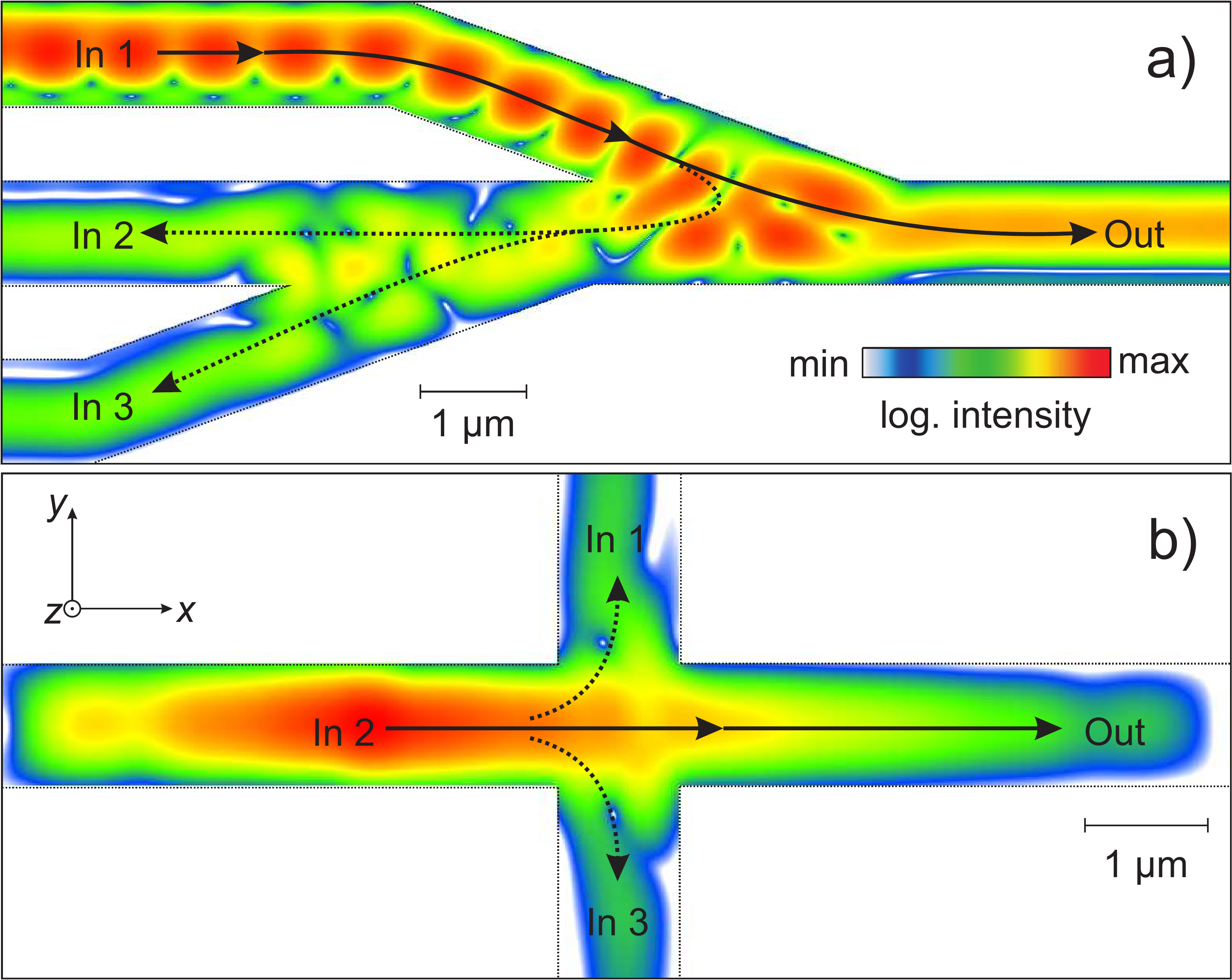}}%
\end{center}%
\caption{\label{fig1} 
Energy transmission for single arm excitations for different majority gate structures. a) For the case $\alpha=20^\circ$ and $x_1 \neq x_2$ a large output signal is obtained from the outer input arms. b) For the case $\alpha=90^\circ$ and $x_1=x_2$ an energy transmission of 72\,\% into the opposed waveguide is achieved.}%
\end{figure}%

To study the performance of the majority gate, a single input arm was excited and the transmission of energy into the output was investigated.
In Fig.~\ref{fig1}a) the energy transmission is shown in a logarithmic colorscale for the spin-wave excitation in input 1 when the merging angle of the waveguide is $\alpha=20\,^\circ$ and the merging position is different for the outer input arms ($x_1\neq x_2$). With this design the energy transmission from input 1 to the output waveguide is 64\,\%, which is the ratio of the maximum value in the output waveguide to the maximum value around $x_2$. The spin waves from input 3 have the smallest transmission of 32\,\% and spin waves from input 2 exhibit an energy transmission of 45\,\% into the output waveguide.

The transmission strongly depends on the bend angle $\alpha$ and the number of merging areas which have to be passed to reach the output. This first point can be understood when the wavevector of the spin waves in the combiner region is split into its components parallel to the $x$- and $y$-direction. With a small bend angle $\alpha$ the $x$-component of the wavevector increases and the $y$-component decreases, so that the spin-wave propagation into the direction of the output waveguide is favored. At the same time, the propagation of the spin waves into the opposed waveguide is suppressed due to the non-overlapping merging areas ($x_1 \neq x_2$). 
The second point becomes clear when the merging areas are considered as source of spin-wave reflections (e.g. in the vicinity of $x_1$ and $x_2$) into the other input arms. This reflection process gives rise to the standing interference patterns visible in Fig.~\ref{fig1}a). The more merging areas are passed, the more reflections can occur.

In Fig.~\ref{fig1}b) another interesting feature of isotropic FVMSW is shown, for the extreme situation when the waveguides are merged under an angle of $90^\circ$ at the same positions ($x_1=x_2$) and when the spin waves are excited in the center waveguide. In this case the wavevector of the incoming spin wave points directly into the output waveguide, and a high energy transmission of 72\,\% through the gate can be achieved, while only 14\,\% of the energy are transmitted to the adjacent input waveguides. This example shows that spin-wave networks can be realized with two-dimensional rectangular crossings in the out-of-plane magnetized geometry. This allows for via-free crossings of spin-wave waveguides, and it is a major advantage of the spin-wave technology, which can examplarily be used in magnonic holographic memories \cite{Gertz2014} or magnonic full adders, based on majority gates.\cite{Ibrahim2008}

\begin{figure}[t]%
\begin{center}%
\scalebox{1}{\includegraphics[width=\linewidth,clip]{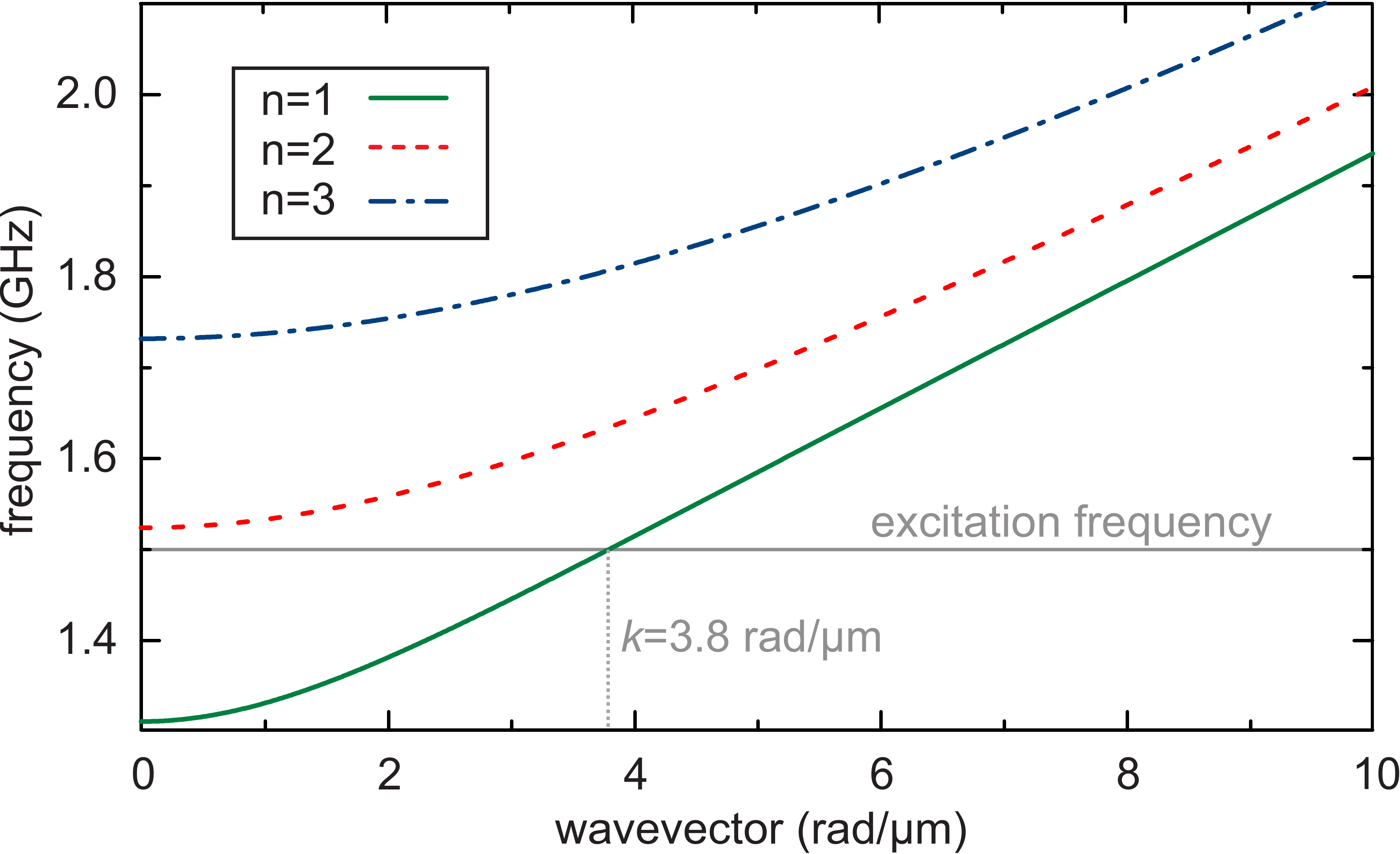}}%
\end{center}%
\caption{\label{disp} 
Dispersion relations for different width modes for a perpendicular magnetized stripe with an effective width of $1.42\,\mu$m. For an excitation frequency of 1.5\,GHz no higher width modes can exist in the waveguide.}%
\end{figure}%

As mentioned above, the interference patterns in Fig.~\ref{fig1}a) result from back reflections in the combiner regions and are not originating from scattering processes into higher dipolar spin waves or exchange spin waves as in the in-plane magnetized gates.\cite{Klingler2014a} This can be understood by examining the dispersion relations\cite{Kalinikos1986} for the first three width modes $n$ of the waveguide which are shown in Fig.~\ref{disp}. Here, $n$ is the number of anti-nodes across the waveguide width. The dispersions were calculated for an effective stripe width of 1.42\,$\mu$m to include dynamic demagnetization effects and effective dipolar pinning.\cite{Guslienko2002} Additionally, a demagnetization factor of 0.92 was used to correct the internal static field in the waveguides.\cite{Joseph1965} It can be seen, that no higher width modes can exist in the waveguides at an excitation frequency $f=1.5$\,GHz. In contrast to the situation in in-plane magnetized waveguides, the direct scattering processes under energy conservation into exchange dominated spin waves are also not possible. Furthermore, one can extract a wavevector of $k=3.8$\,rad/$\mu$m, a wavelength of $\lambda=1.7\,\mu$m and a group velocity of $v_\text{g}=0.13\,\mu$m/ns from the dispersion relation at the excitation frequency $f$. Together with the decay time of the spin wave of $\tau=566$\,ns, which can be calculated from the material parameters,\cite{Patton1975} one obtains a decay length $\lambda_\text{dec}=v_\text{g}\tau=73\,\mu$m, which is much larger than the size of the microstructures.

For the majority operation it is important to account for the different propagation losses and phase shifts resulting from the different propagation distances. For this, we compare the output phase and amplitude of all single arm excitations of the gate in Fig.~\ref{fig1}a). As a result we reveal as equalizing parameters for input 2 an attenuation factor of 0.7 and a phase shift of $\Delta \phi = -0.35\pi$ with respect to input 3. The spin-wave excitation in input 1 has to be attenuated by a factor of 0.5 and the excitation phase has to be shifted by $\Delta \phi = -0.64\pi$ relative to input 3. The different equalizing factors can be understood by the asymmetric gate geometry. In a real device these shifts can be realized by a small displacement of the antenna positions or by the use of nanomagnets as phase shifters.\cite{Au2012}
With this adjustment parameters the output signals of every single input coincide, and {all possible} input combinations can be simulated.{\cite{Klingler2014a}}

\begin{figure}[t!]%
\begin{center}%
\scalebox{1}{\includegraphics[width=\linewidth,clip]{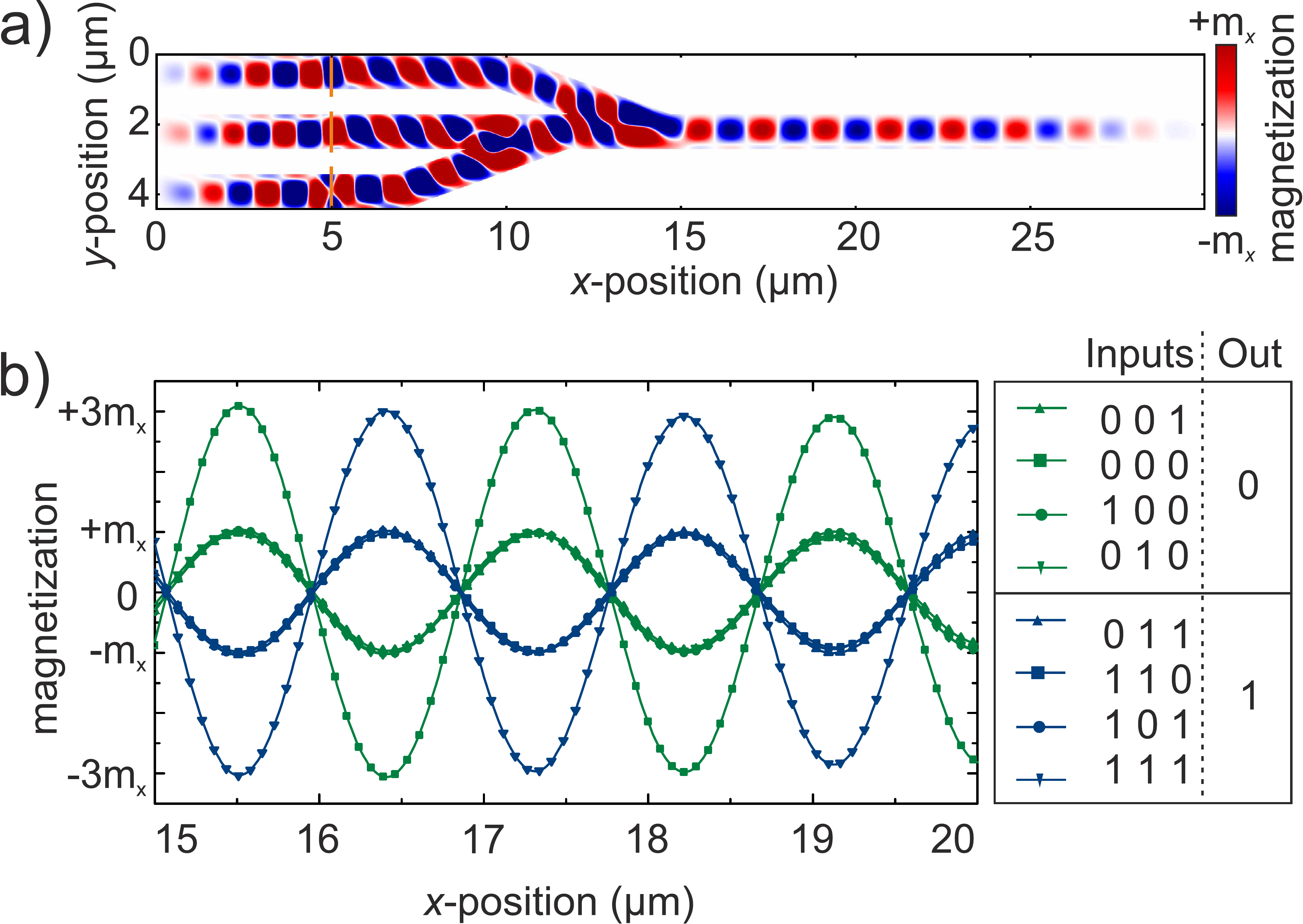}}%
\end{center}%
\caption{\label{fig3} 
a) Spatial magnetization distribution of a fixed time step for the 1-0-0 operation. The color code shows the deviation in the magnetization due to the spin wave propagation. b) All input combinations are shown in one graph. Input combinations with the same majority phase are in phase. Input combinations with different majority phase are out of phase.}%
\end{figure}%

In Fig.~\ref{fig3}a) the $x$-component of the magnetization distribution $m_x$ is shown for a logic 1-0-0 excitation (logic values from top to bottom) at a fixed time step. This means, that the spin waves in the upper input are excited with an additional phase shift of $\pi$, whereas the spin waves in the center and bottom waveguide are excited without an additional phase shift. At the edge of the waveguides one can see a small area where the magnetization is inverted. This is the result of an overlapping edge-mode\cite{Damon1961,Lisenkov2014} which propagates on the edge of the waveguide and decays exponentially into the waveguide width. Due to its small amplitude, no negative influences of this effect are expected.

In Fig.~\ref{fig3}b) $m_x$ of every possible input combination is shown for the same time step. The sinusoidal shape of the magnetization distributions is clearly visible. No disturbing influences of exchange waves can be seen, compared to in-plane magnetized majority gates.\cite{Klingler2014a} All spin-waves with a majority phase of ``0'' are in phase and are exactly out of phase to the spin waves with a majority phase of ``1''. This is the clear signature of the majority function, which can be seen by comparing the results with the truth table in the legend in Fig.~\ref{fig3}b). From the magnetization distributions one can easily extract a wavelength of about 1.8\,$\mu$m, which is in good agreement with the prediction of the dispersion relation. Additionally, the amplitudes of the spin waves in the output waveguide fit very well to the theoretical predictions: if all three input waves are in phase, the amplitude in the output waveguide is three times larger than for every other case. This confirms the adjustment parameters in the input waveguides. Since the information is encoded in the spin-wave phase, the different output amplitudes are not essential for the information content of the spin wave. However, if the output signal should be forwarded to another logic gate, the amplitude has to be normalized before, e.g. by the use of parametric amplification \cite{Serga2003,Khitun2009,Bracher2014} or by use of a magnon transistor.\cite{Chumak2014}

{To show that the spin-wave majority gate allows for a full set of logic operations, input 3 can be chosen as a control input. At a fixed readout position, e.g. at $x=15.5\,\mu$m, one can perform AND-operations between inputs 1 and 2 when the spin waves in input 3 are excited without any additional phase shift. At the same readout position OR-operations can be performed when the spin waves in input 3 are excited with an additional phase shift of $\pi$. By shifting the readout position by one half of the wavelength, e.g. to $16.4\,\mu$m, the output phase of the spin wave is shifted by $\pi$ and thus, \mbox{NAND-} and NOR-operations are performed dependent on the control input signal. In total, four different logic operations can be processed with a single majority gate.}

In conclusion, a fully functional, asymmetric magnonic majority gate design for the use with isotropic FVMSW has been presented. It was shown, that the output signal is not disturbed by parasitic scattering processes into exchange spin waves.
This results in a very clear and undistorted interference pattern in the output waveguide. It was verified, that the majority of the input values defines the output signal, which is a clear signature of a working majority gate. {Furthermore, the spin-wave logic approach was used to perform AND-, OR-, NAND-, and NOR-operations with a single gate structure.} {The data} processing and transmission occured in the pure spin-wave system, which makes the gate suitable for an integration into complex magnonic networks with various daisy-chained spin-wave devices.\cite{Chumak2014, Bracher2013, Vogt2014} {Finally, it was shown that the utilization of isotropic FVMSW allows for the realization of orthogonal waveguide crossings, where spin waves can pass the crossing region with a high transmission without the need of a via-implementation which would demand for several layers and patterning steps. Such spin-wave waveguide crossings are necessary for  spin-wave logic device networks.\cite{Gertz2014,Ibrahim2008}}
 
This research has been supported by the EU-FET grant InSpin 612759.

\end{document}